\documentclass[aps, pra, superscriptaddress, reprint, twocolumn]{revtex4-2}
\usepackage[T1]{fontenc}
\usepackage[utf8]{inputenc}
\usepackage{microtype}
\usepackage{graphicx}
\usepackage{amsthm,amsfonts,amstext,amssymb,amsmath}
\usepackage{latexsym}
\usepackage{braket}
\usepackage{bbold}
\usepackage{bm}
\usepackage{placeins}
\usepackage{enumerate}
\usepackage{standalone}
\usepackage{float}
\usepackage{mathtools}
\usepackage{xcolor}
\usepackage[colorlinks=true,
  linkcolor=black,
  citecolor=blue,
  urlcolor =blue]{hyperref}

\newcommand{\rtext}[1]{\textcolor{red}{ $\to$#1}}
\renewcommand{\rtext}[1]{}

\definecolor{RED}{rgb}{1,0,0}

\newtheorem{theorem}{Theorem}

\DeclareMathOperator{\trace}{tr}
\newcommand*{\tr}[2][]{\ensuremath{{\trace_{#1}}{\left(#2\right)}}}

\begin{document}

\author{Andreas Elben}
\affiliation{Institute for Quantum Information and Matter, Caltech, Pasadena, CA, USA}
\affiliation{Walter Burke Institute for Theoretical Physics, Caltech, Pasadena, CA, USA}
\affiliation{Institute for Theoretical Physics, University of Innsbruck, Innsbruck A-6020, Austria}
\affiliation{Institute for Quantum Optics and Quantum Information of the Austrian Academy of Sciences,  Innsbruck A-6020, Austria}

\author{Steven T. Flammia}
\affiliation{AWS  Center  for  Quantum  Computing,  Pasadena,  CA, USA}
\affiliation{Institute for Quantum Information and Matter, Caltech, Pasadena, CA, USA}

\author{Hsin-Yuan Huang}
\affiliation{Institute for Quantum Information and Matter, Caltech, Pasadena, CA, USA}
\affiliation{Department of Computing and Mathematical Sciences, Caltech, Pasadena, CA, USA}

\author{Richard~Kueng}
\affiliation{Institute for Integrated Circuits, Johannes Kepler University Linz, Altenbergerstrasse 69, 4040 Linz, Austria}

\author{John Preskill}
\affiliation{Institute for Quantum Information and Matter, Caltech, Pasadena, CA, USA}
\affiliation{Walter Burke Institute for Theoretical Physics, Caltech, Pasadena, CA, USA}
\affiliation{Department of Computing and Mathematical Sciences, Caltech, Pasadena, CA, USA}
\affiliation{AWS  Center  for  Quantum  Computing,  Pasadena,  CA, USA}

\author{Beno\^it Vermersch}
\affiliation{Institute for Theoretical Physics, University of Innsbruck, Innsbruck A-6020, Austria}
\affiliation{Institute for Quantum Optics and Quantum Information of the Austrian Academy of Sciences,  Innsbruck A-6020, Austria}
\affiliation{Univ.\ Grenoble Alpes, CNRS, LPMMC, 38000 Grenoble, France}

\author{Peter Zoller}
\affiliation{Institute for Theoretical Physics, University of Innsbruck, Innsbruck A-6020, Austria}
\affiliation{Institute for Quantum Optics and Quantum Information of the Austrian Academy of Sciences,  Innsbruck A-6020, Austria}

\begin{abstract}
   Increasingly sophisticated programmable quantum simulators and quantum computers are opening unprecedented opportunities for exploring and exploiting the properties of highly entangled complex  quantum systems. The complexity of large quantum systems is the source of their power, but also makes them difficult to control precisely or characterize accurately using measured classical data. We review recently developed protocols for probing the properties of complex many-qubit systems using measurement schemes that are practical using today’s quantum platforms. In all these protocols, a quantum state is repeatedly prepared and measured in a randomly chosen basis; then a classical computer processes the measurement outcomes to estimate the desired property. The randomization of the measurement procedure has distinct advantages; for example, a single data set can be employed multiple times to pursue a variety of applications, and imperfections in the measurements are mapped to a simplified noise model that can more easily be  mitigated. We discuss a range of use cases that have already been realized in quantum devices, including Hamiltonian simulation tasks, probes of quantum chaos, measurements of nonlocal order parameters, and comparison of quantum states produced in distantly separated laboratories. By providing a workable method for translating a complex quantum state into a succinct classical representation that preserves a rich variety of relevant physical properties, the randomized measurement toolbox strengthens our ability to grasp and control the quantum world. 
\end{abstract}

\title{The randomized measurement toolbox}

\maketitle

\section{Introduction and Motivation}

As far as we know, it is not possible using classical data to fully and succinctly characterize generic quantum systems of many strongly interacting particles. This observation is both a curse and a blessing. On the one hand, it limits the ability of classical beings like us to grasp the behavior of complex highly entangled quantum systems. On the other hand, it invites us to build and operate large-scale quantum systems that can perform useful tasks beyond what we can imagine.

The emergence of increasingly powerful quantum technologies has transformed the challenge of characterizing complex quantum systems from a theoretical conundrum to a laboratory imperative. Someday we will have large-scale error-corrected quantum computers to help us advance the frontiers of quantum physical science and run useful applications. 
While these dream machines may still lie far in the future, even today highly programmable quantum platforms \cite{preskill2018noisy,altman2021quantum} can create and control complex states comprising many atoms \cite{gross2017quantum,schaefer2020tools,browaeys2020many,morgado2021quantum,blatt2012quantum,monroe2021programmable}, spins \cite{kloeffel2013prospects,burkard2021semiconductor}, photons \cite{slussarenko2019photonic,pelucchi2021potential}, or superconducting circuit elements \cite{kjaergaard2020}, opening unprecedented opportunities for scientific discovery \cite{alexeev2021quantum}. 

Experimentalists and theorists working together must develop, perfect, and employ suitable tools to investigate and exploit the features of many-qubit quantum states that are created in the laboratory. This typically involves preparing and measuring the same quantum state over and over again. With sufficiently many repetitions, it is possible to completely characterize an $n$-qubit state by means of full state tomography, but this task is hopelessly inefficient, requiring a number of experiments exponential in $n$ \cite{flammia2012compressed-sensing,haah2016tomography,odonnel2016tomography}, and an amount of classical postprocessing of the experimental results which is also exponentially large. Fortunately, a far less complete description of the state is adequate for many purposes \cite{aaronson2018shadow,aaronson2019gentle}, so that the number of experiments and the amount of classical processing needed can be drastically reduced. In this article, we review some recent theoretical ideas about how to improve the efficiency of characterizing complex quantum states, and some of the experimental results that flow from these ideas.

The concepts and examples we discuss here share a common theme. 
Rather than tailoring the measurements we perform in the lab to the particular properties we wish to study, we can instead repeatedly perform measurements which are randomly sampled from a fixed ensemble, and then adapt the classical postprocessing of the measurement outcomes to the particular task at hand \cite{vanenk2012measuring,elben2018renyi,elben2019statistical,huang2020predicting,paini2019,morris2019selective,knips2020multipartite,ketterer2019characterizing}. 
This randomized measurement strategy can be surprisingly powerful even when the measurements are simple enough to be performed with adequate precision using today’s noisy quantum platforms. 
A particularly simple procedure is to measure each qubit in a randomly chosen basis. 
By repeating this procedure of order $\log (L)$ times, and using only efficient classical postprocessing, we can accurately estimate the expectation values of any $L$ local operators --- the number of experiments needed does not depend at all on the total number of qubits \cite{huang2020predicting}. 
Randomized single-qubit measurements also enable us to estimate properties of larger subsystems \cite{elben2018renyi}; in this case the cost rises exponentially with the size of the subsystem, but is still far lower than the cost of complete tomography of the subsystem \cite{elben2018renyi,elben2019statistical,huang2020predicting}. 
Alternatively, global properties of the quantum state can be estimated using a modest number of measurement repetitions if the measurements are preceded by relatively efficient information scrambling unitary operations executed using a quantum computer or programmable quantum simulator \cite{huang2020predicting}. 
A further advantage of the randomized measurement approach is that randomization simplifies the effects of noise, so that imperfections in measurement outcomes can be more easily mitigated by suitably modifying the classical postprocessing of the outcomes \cite{chen2020robust,koh2020classical}. 

Many applications of this randomized measurement toolbox have already been conceived and executed in experiments using quantum devices. 
We can estimate the overlap of two quantum states produced in separate laboratories far apart from one another~\cite{elben2020cross,zhu2021crossplatform}. 
We can probe chaotic quantum dynamics by measuring out-of-time-order correlation functions~\cite{vermersch2019probing,joshi2020quantum}, without reversing time evolution or introducing ancilla systems. 
We can quantify quantum entanglement by measuring entropy~\cite{brydges2019probing} and other entanglement measures~\cite{elben2020mixed}. 
We can compute order parameters that characterize topological order or symmetry-protected topological order~\cite{elben2020many}. 
We can estimate the expectation value and variance of a local Hamiltonian~\cite{huang2020predicting}. 

These and other applications have a notable feature. 
First repeated randomized measurements map a multi-qubit quantum state to succinct classical data. 
Later, these classical data are processed to investigate properties of interest. 
Conveniently, the properties to be investigated need not be known when the measurements are performed using the quantum device; rather we can:
\begin{center}
    \textit{Measure first, ask questions later.}
\end{center}
Indeed, some of the applications we review were carried out by reanalyzing data that had originally been taken with a different purpose in mind. 

A randomized measurement protocol may be viewed as a feasible scheme for translating the extravagant quantum information residing in a many-qubit state into a succinct classical representation of the state. This quantum-to-classical conversion process unavoidably discards a vast amount of information about the state, but the burgeoning applications illustrate that many physically relevant features of the state can survive. Thus scientists assisted by their powerful classical computers, by pondering and manipulating classical data, can grasp crucial properties of the quantum world that might otherwise remain concealed.

In Section~\ref{sec:exp_rec} we review a particular randomized measurement scheme using repeated single-qubit measurements to construct a ``classical shadow’’ of a quantum state, state a rigorous guarantee on the accuracy of estimated operator expectation values based on classical shadows, and describe an application of randomized single-qubit measurements  to estimating the purity of a many-qubit state. Section~\ref{sec:oneidea} surveys a variety of other use cases for randomized measurements, including applications to Hamiltonian simulation tasks, variational quantum algorithms, classifying quantum phases of matter, fidelity estimation, noise characterization, and cross-platform comparison of quantum systems. While all these applications target many-qubit systems, Section~\ref{sec:challenges} addresses extensions to higher-dimensional qudits and to systems of bosons and fermions. We also discuss robustness of randomized measurement protocols against noise, and comment on experimental protocols that do not require local control of the quantum system. We conclude by noting that classical representations derived from randomized quantum-to-classical conversion schemes, when combined with classical machine learning tools, provide leverage for predicting the behavior of quantum systems beyond those already encountered in the laboratory.

\section{Experimental recipe and postprocessing of the measurements }
\label{sec:exp_rec}

\begin{figure}[t]
    \centering
    \includegraphics[width=0.99\linewidth]{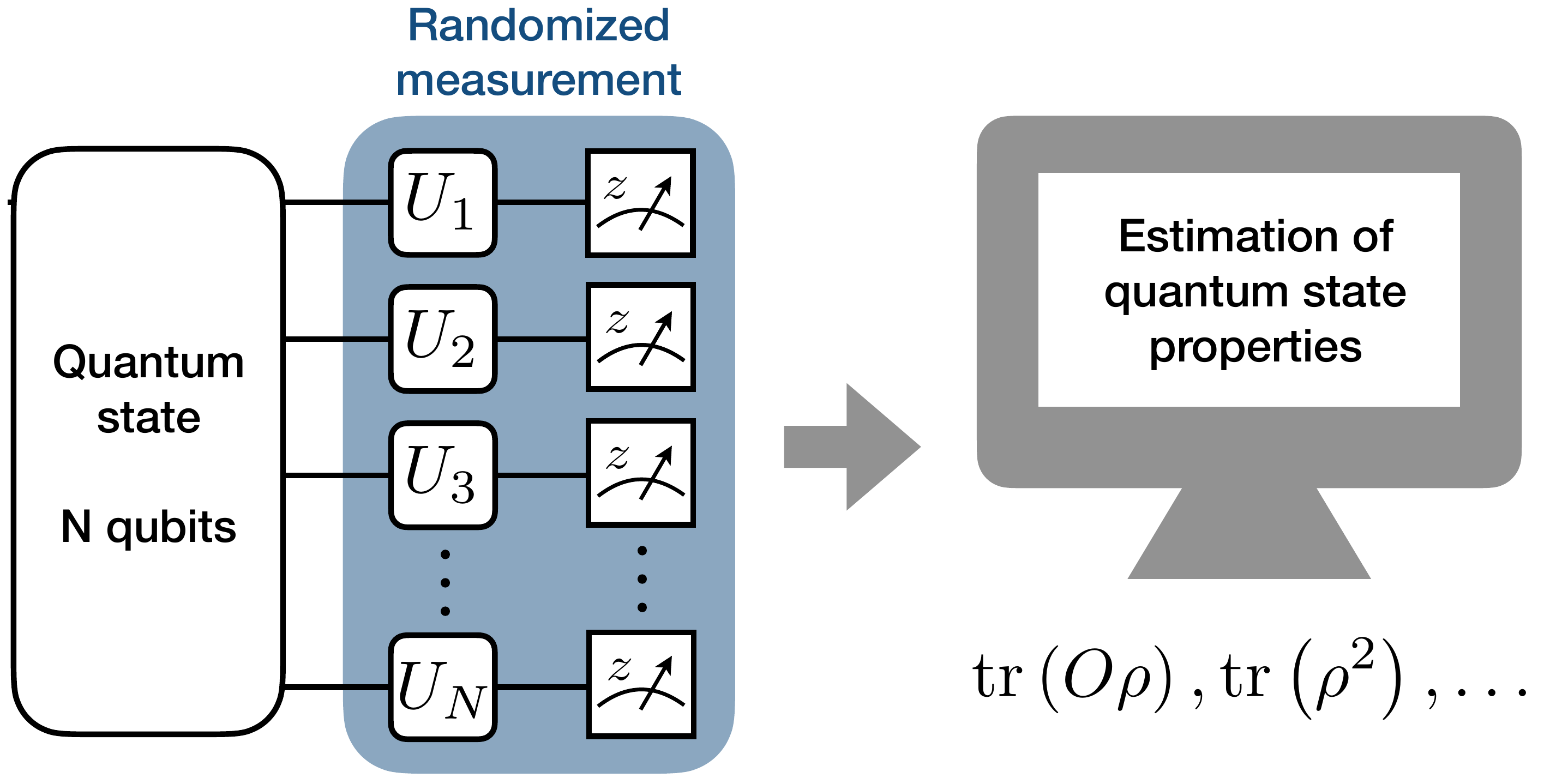}
    \caption{\textit{Randomized measurements} are implemented on a $N$ qubit quantum state $\rho $ via the application of local random unitaries $U=\bigotimes_{n=1}^N U_i$   and subsequent projective measurements performed in the computational $z$-basis. Via classical post-processing of the outcomes, many properties of $\rho$, such as observable expectation values $\tr{O\rho}$ and the purity $\tr{\rho^2}$, can be estimated. Remarkably, this estimation is \emph{provably efficient} if we restrict attention to properties of (arbitrary) subsystems.}
    \label{fig:random_meas_scheme}
\end{figure}

\subsection{Protocol---Data acquisition}
\label{sec:pro_dat}

We consider a quantum system consisting of $N$ qubits with associated Hilbert space \mbox{$(\mathbb{C}^2)^{\otimes N}$}. A \textit{randomized measurement} (RM) is schematically presented in Fig.~\ref{fig:random_meas_scheme} and consists in the following steps. $(i)$ The quantum many-body state $\rho$ of interest is prepared in the device. $(ii)$ A unitary operation $U$, selected at random from a suitable ensemble of unitary operations, is applied to $\rho$.
For concreteness, we consider here local random operations $U=\bigotimes_{n=1}^N U_n$, applied to each qubit independently.  The individual single-qubit rotations $U_n$ ($n=1,\dots,N$) are sampled from ensembles of single-qubit unitary operations which evenly cover the Bloch sphere of each qubit. 
Examples of such unitary designs \cite{gross2007evenly,dankert2009exact} include the single-qubit Clifford group, as well as the full unitary group $U(2)$ encompassing all single-qubit transformations. 
$(iii)$ Lastly, a projective measurement in the computational basis $\{\ket{\mathbf{s}}\}$ is performed, with outcome bit string $\mathbf{s}=(s_1, \dots,s_N)$ and $s_n \in \{0,1\}$ for $n=1,\dots,N$.
Steps $(i)$-$(iii)$ are then repeated $K$ times with a fixed unitary $U$. 
Subsequently, the entire procedure is repeated with $M$ independently sampled unitaries $U$ such that in total $M\cdot K$ experimental runs are performed.

In summary, $M \cdot K$ experimental runs are executed, each of which is characterized by $N$ single-qubit unitaries $U_1^{(m)},\ldots,U_N^{(m)}$ that only depend on $m$, and an $N$-bit outcome $\mathbf{s}^{(m,k)}=(s_1^{(m,k)},\ldots,s_N^{(m,k)})$ that depends on both $m$ and $k$. Storing both up to floating point accuracy is comparatively cheap, requiring storage of $\mathcal{O}(MKN)$ floating point numbers in total.

\subsection{Protocol---Postprocessing }
\label{sub:postprocessing}

After completing a full experiment, 
we can use 
the obtained data
to extract information about the underlying many-body system. It is instructive to consider two extreme examples. For the sake of simplicity, we also assume that the single-qubit rotations $U_i$ are sampled from 
the discrete ensemble that randomly permutes Pauli matrices  $\left\{X,Y,Z\right\}$ (the Clifford group). That is, $U_n Z U_n^\dagger = W_n \in \left\{X,Y,Z\right\}$ for $1 \leq n \leq N$.

\begin{figure}[t]
    \centering
    \includegraphics[width=0.84\linewidth]{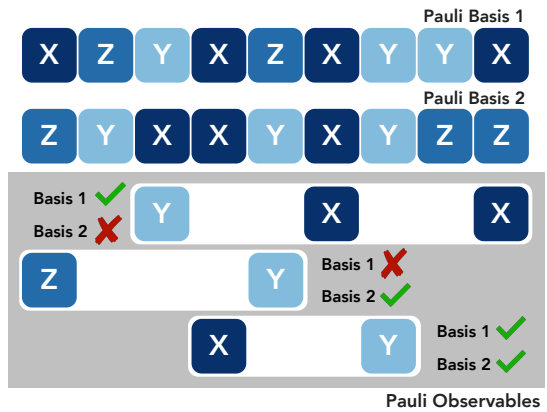}
    \caption{Given a Pauli string that denotes the basis one measures in for each qubit. A Pauli observable $O$ given by a tensor product of $\{I, X, Y, Z\}$ is compatible with the string if the non-identity part of $O$ matches the string.}
    \label{fig:pauli-compati}
\end{figure}

Setting $M=1$ means that we repeat the same randomized measurement over and over ($K>1$ times). This is equivalent to measuring a random string of Pauli observables, namely $U_1^\dagger Z U_1 \otimes \cdots \otimes U_n^\dagger Z U_N = W_1 \otimes \cdots \otimes W_N$, a total of $K$ times. This, in turn, allows us to approximate the expectation value $\mathrm{tr} \left(W_1 \otimes \cdots \otimes W_N \rho \right)$
as well as compatible subsystem marginals,
e.g.\ $\mathrm{tr} \left( W_1 \otimes \cdots \otimes  W_l \otimes \mathbb{I}^{\otimes (N-l)} \rho \right)$ for $1 \leq l \leq N$ (see Fig.~\ref{fig:pauli-compati}).
Other Pauli expectation values are off limits, though.

The other extreme case pilots us into more interesting territory. 
We sample a total of $M$ random Pauli strings $W_1^{(m)} \otimes \cdots \otimes W_N^{(m)}$, $1 \leq m \leq M$ and measure each of them exactly once, $K=1$. A single measurement outcome does not permit reliable approximation of any of the original Pauli expectation values. But, we can combine samples across different measurements to predict many (subsystem) expectation values. Take $X \otimes Y \otimes Z \otimes \mathbb{I}^{\otimes (N-3)}$ as a concrete example. 
The outcome from measuring $W_1^{(m)} \otimes \cdots \otimes W_N^{(m)}$ provides useful statistical information if and only if $W_1^{(m)} =X$, $W_2^{(m)} = Y$ and $W_3^{(m)}=Z$. If we assign all single-qubit unitaries $U_n$ uniformly at random, these accordances occur with probability $(1/3)^3$. In turn, we can expect that a total of $M \geq 3^3/\epsilon^2$ randomly selected $N$-qubit Pauli measurements provide enough statistical data to $\epsilon$-approximate $\mathrm{tr} \left( X \otimes Y \otimes Z \otimes \mathbb{I}^{\otimes (n-3)} \rho \right)$. Interestingly, this argument only depends on subsystem size, $w=3$ for our example, while the actual qubit locations and Pauli strings of interest ($X \otimes Y \otimes Z \otimes \mathbb{I}^{\otimes (N-3)}$ in our example) do not matter  at all. In fact, as summarized in Theorem 1 below, we can extend our argument to cover (very) many size-$w$ expectation values in one go.

The actual prediction step is also relatively straightforward. We restrict our attention to measurement settings that are compatible with the Pauli expectation value $o=\mathrm{tr}(O \rho)$ in question and use empirical averaging of compatible outcomes to obtain an approximation $\hat{o}$  of $o$. 
The following formula succinctly summarizes such an estimation process:
\begin{align}
\hat{o} =& \frac{1}{M}\sum_{m=1}^M\mathrm{tr}(O  \hat{\rho}^{(m)}) \quad \mathrm{where} \label{eq:linear-prediction} \\
\hat{\rho}^{(m)}=&  \frac{1}{K }\sum_{k=1}^K \, \bigotimes_{n=1}^N \left( 3 (U_n^{(m)})^\dagger |s_n^{(m,k)} \rangle \! \langle s_n^{(m,k)}| U_n^{(m)}-\mathbb{I} \right) \label{eq:classical-shadow}
\end{align}
combines the $m$-th measurement settings ($U_1^{(m)},\ldots,U_N^{(m)}$) with the $K$ associated  outcomes ($s^{(m,k)}_1,\ldots,s^{(m,k)}_N$) to produce an approximation of the underlying $N$-qubit quantum state $\rho$. The collection $\{ \hat{\rho}^{(m)} \}_{m=1,\dots,M}$  is  called a \emph{classical shadow} of $\rho$ \cite{huang2020predicting,paini2019}. 
This procedure works for arbitrary target observables $O$ (not just Pauli expectation values), and ensembles of single-qubit random unitaries which cover the Bloch sphere evenly (not just Clifford unitaries).

Remarkably, randomized measurements  give access not only to observables, 
but also polynomial functionals of the density matrix.
In fact, RMs were first
envisioned to estimate 
the \emph{purity}
$P_2 = \mathrm{tr}(\rho^2)$ 
by means of the following formula \cite{vanenk2012measuring,elben2018renyi,brydges2019probing,elben2019statistical}
\begin{eqnarray}
\hat P_2&=&
\frac{2^N}{M K (K-1)} \sum_{m=1}^M \sum_{\substack{k,k'=1 \\ k\neq k'}}^K  (-2)^{-D[\mathbf{s}^{(m,k)},\mathbf{s}^{(m,k')}]}. \label{eq:puritybitstrings}
\end{eqnarray}
We calculate first a weighted average of Hamming distances $D\big[\mathbf{s}^{(m,k)},\mathbf{s}^{(m,k')}\big]$ of two distinct outcomes $\mathbf{s}^{(m,k)}$ and $\mathbf{s}^{(m,k')}$ that belong to the same measurement setting $m$. Subsequently, we average over different measurement settings. While the precise form of Eq.~\eqref{eq:puritybitstrings} follows from derivations presented in Refs.~\cite{brydges2019probing,elben2019statistical}, some intuition might be gained as follows: We observe bitstrings $\mathbf{s}^{(m,k)}$ according to their Born probabilities $P_U(\mathbf{s}^{(m,k)})=|\braket{\mathbf{s}^{(m,k)}| U^{(m),\dagger} \rho   U^{(m)} | \mathbf{s}^{(m,k)}   }|^2 $. In Eq.~\eqref{eq:puritybitstrings},  second order correlations of (estimations of) these Born probabilities are averaged over local random unitaries. Such an average must correspond to a second order functional of $\rho$ which is invariant under local random unitary transformations, i.e., in our case, the purity.

Alternatively, one can estimate the purity by replacing distinct copies of $\rho$ by distinct classical shadows~\eqref{eq:classical-shadow} and average over all possible choices \cite{huang2020predicting,elben2020mixed}:
\begin{equation}
\hat{P}_2 = \frac{1}{M(M-1)}\sum_{m \neq m'}
\mathrm{tr} \left( \hat{\rho}^{(m)} \hat{\rho}^{(m')}\right). \label{eq:shadow-purity}
\end{equation}
This estimation procedure also extends to arbitrary polynomials of the density matrix \cite{huang2020predicting,elben2020mixed,vitale2021symmetry,rath2021quantum}. 

With both estimators, we can access purities $\tr{\rho_A^2}$ of reduced density matrices $\rho_A=\tr[A^c]{\rho}$ of arbitrary subsystems $A$ (with complement $A^c$) by restriction  during the postprocessing.
For a fixed total number of experimental runs $MK$, Eq.~\eqref{eq:puritybitstrings} 
achieves a more accurate estimate for many repetitions $K$ of a few measurement settings $M$, while  Eq.~\eqref{eq:shadow-purity} performs better for 
many different measurement settings with few repetitions each. 
In addition, the estimator Eq.~\eqref{eq:puritybitstrings} is expected to be more robust against miscalibration of the random unitaries compared to Eq.~\eqref{eq:shadow-purity}. This is due to the fact that the estimator in Eq.~\eqref{eq:puritybitstrings} depends solely on the measured bitstrings and matrix elements of the applied random unitaries do not appear explicitly; see also Sec.~\ref{sec:vignette}.

\subsection{Rigorous theory and history}

The postprocessing rules introduced in Eqs.~\eqref{eq:linear-prediction}, \eqref{eq:classical-shadow} can be equipped with rigorous error bounds. 
Here, we present an exemplary performance guarantee that is valid for evenly distributed ensembles of single-qubit unitaries, like the full unitary group or the Clifford group. 

\begin{theorem} \label{thm:linear-error-bound}
$M \propto \log (L) 4^{w} /\epsilon^2$ independent randomized measurements suffice to $\epsilon$-approximate an entire collection of $L$ subsystem-size-$w$ expectation values with high success probability.
\end{theorem}

For the special case of Pauli expectation values, an improved scaling of $M \propto \log(L) 3^w/\epsilon^2$ randomized measurements readily follows from the arguments provided in Sub.~\ref{sub:postprocessing}.
Historically, this result for Paulis predates Theorem~\ref{thm:linear-error-bound}, and was first proven in Ref.~\cite[Appendix~D]{evans2019scalable} by a slightly different but equally simple argument to the one presented here. 
This result in turn was influenced by the earlier, but quadratically weaker bound in Ref.~\cite{cotler2020quantum}. 
The general case displayed in Theorem~\ref{thm:linear-error-bound} is based on
the arguments presented in Ref.~\cite{huang2020predicting}.
The actual error bound implicitly works in the single-shot limit ($K=1$). 
But, multiple repetitions for each measurement setting ($K >1$) can only further improve performance. 

Theorem~\ref{thm:linear-error-bound} contains an interesting tradeoff between subsystem size (which enters exponentially) and the number of observables (which enters logarithmically). 
For instance, already  
$M \propto \log (N)/\epsilon^2$ randomized $N$-qubit Pauli measurements suffice to $\epsilon$-approximate \emph{all} 2-body Pauli expectation values in a system with $N$ qubits. 
And, remarkably, the statement is completely independent of the underlying quantum state $\rho$.

This showcases that it can be much easier (and more reliable) to accurately approximate certain properties of an unknown state than estimating the full state $\rho$. 
This related problem, called \emph{quantum state tomography}, has a long and prominent history \cite{gross2013tomography}. 
Fundamental lower bounds assert that $\epsilon$-accurate quantum state tomography of an $N$-qubit system must require exponentially many samples in general ($M \cdot K \geq 4^N/\epsilon^2$, see \cite{haah2017sample,O'Donnell2016Jun,O'Donnell2017Jun}).
\textcolor{black}{Substantial improvements are only possible if the state in question is known to have (very) advantageous structure, e.g.~matrix product states with polynomial bond dimension \cite{cramer2010efficient} or neural network states \cite{torlai2018neural}.}
Regarding the measurement of the purity, for both estimators Eqs.~\eqref{eq:puritybitstrings} and \eqref{eq:shadow-purity}, the required number of experimental runs $MK$ to obtain a given accuracy scales exponentially with system size $N$ \cite{huang2020predicting,elben2020mixed,rath2021importance}, but with a significantly reduced exponent compared to full state tomography.

The idea of bypassing quantum state tomography, i.e.\ the full reconstruction of the quantum state $\rho$, and directly predicting  (subsystem) expectation values  $\tr{O\rho}$ is also known as \emph{shadow estimation} \cite{aaronson2018shadow,aaronson2019gentle}. 
In its original form, shadow estimation does not have an exponential dependence on subsystem size $w$, but does require loading multiple state copies into a quantum memory and performing entangling quantum computations on them. 
The procedure discussed here can be viewed as a more near-term variant of this idea. 
But it also draws inspiration from Refs.~\cite{vanenk2012measuring,ohliger2013efficient,morris2019selective}, as well as resource-efficient approaches to quantum state tomography  \cite{sugiyama2013precision,guta2020fast}.

\subsection{Vignette application: Purity measurements in an ion trap quantum simulator\label{sec:vignette}}
\begin{figure}[t]
    \centering
    \includegraphics[width=1\columnwidth]{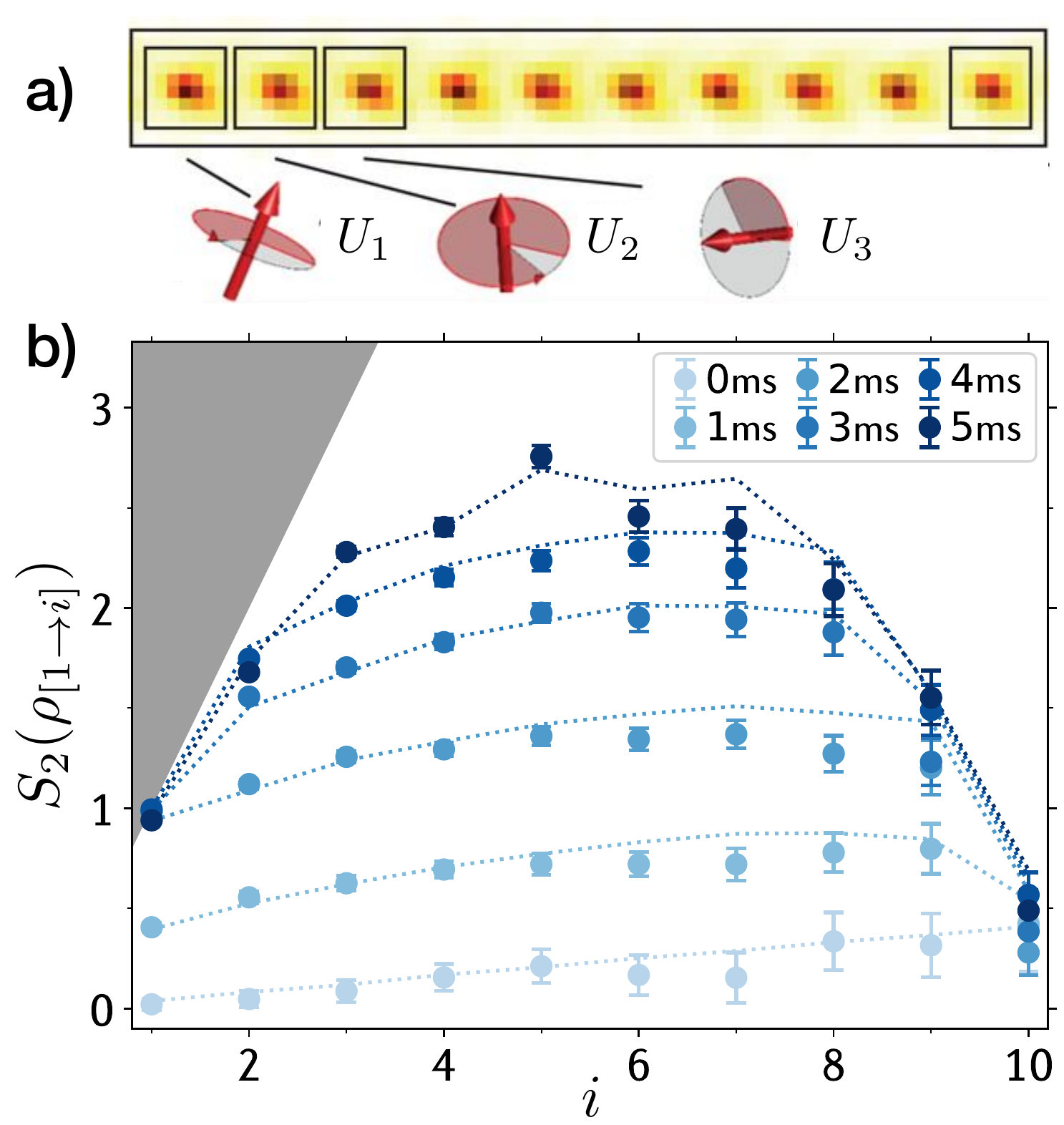}
    \caption{
    \textit{Estimation of the second R\'{e}nyi entropy with RMs.}
    a) In Ref.~\cite{brydges2019probing}, randomized measurements have been implemented in a 10-qubit trapped ion system utilizing single-qubit Haar random unitaries followed by a computational basis measurement. In this case, the unitaries $U_i=R_i^X(-\pi/2) R_i^Z(\alpha_i) R_i^X(\pi/2) R_i^Z(\beta_i)$  can be decomposed into uniform $X$-rotations $R_i^X(\pm \pi/2) = \exp(\mp i X_i \pi/4) $ and local $Z$-rotations $R^Z_i(\alpha_i)= \exp(-\mathrm{i} Z_i \alpha_i/2)$. 
    b) Experimental results from Ref.~\cite{brydges2019probing} for the second R\'{e}nyi  $S_2(\rho_A)=-\log_2(P_2)$ of partitions $A=[1,2, \dots, i]$ ($i=1,\dots, 10$) in a system with $N=10$ ions in total. Colored dots correspond to different evolution times $t=0,\dots,5$ ms, with  error bars denoting the standard error of the mean. Dotted lines display  results of  numerical simulation including decoherence effects.
    }
    \label{fig:purity_exp}
\end{figure}

Let us take Ref.~\cite{brydges2019probing} as an illustrative example of RMs.
The goal was to measure the purity in an ion-trap quantum simulator. 
This is relevant for checking that a quantum device works as intended, i.e.\ that the realized quantum state is pure. 
Subsystem purities can also be used to quantify entanglement within quantum many-body systems in terms of the second R\'enyi entropy $S_2(\rho_A)=-\log_2(P_2)$ \cite{eisert2010colloquium}.
Trapped ion quantum simulators contain an array of ions, $N=10$ or $N=20$ in this case, and each encodes a qubit using two long-lived electronic states. 
These can then be manipulated using focused laser beams. 
The system was propagated from an initial N\'eel state $\ket{\psi}=\ket{01}^{\otimes (N/2)}$ to an entangled state $\ket{\psi(t)}=e^{-\mathrm{i}H_{XY}t}\ket{\psi}$ using a Hamiltonian
$H_{XY}=\sum_{i<j} J_{ij} (\sigma_i^+ \sigma_j^-+\mathrm{h.c.}),$
with $J_{ij}\sim J/|i-j|^\alpha$, and $0<\alpha<3$. Due to  preparation errors, dephasing, spontaneous emission, etc, the system after the evolution time $t$ is described by a density matrix $\rho(t)$.
Randomized measurements were implemented by sampling individual single-qubit rotations from the circular unitary ensemble and decomposing them into rotations along the $z$ and $x$ axes. This process is illustrated in Fig.~\ref{fig:purity_exp}a.
Importantly, it only requires local $z$-rotations, while the ions were rotated along the $x$-axis via a global beam.
The total data acquisition involved $M=500$ RM settings with $K=150$ single-shot repetitions each. Postprocessing was based on the averaged purity formula~\eqref{eq:puritybitstrings} which is designed to process many repetitions per measurement setting ($K \gg 1$). 

The plot in Fig.~\ref{fig:purity_exp}b highlights that this RM protocol faithfully estimates second R\'enyi entropies for a variety of different subsystems $A$ and evolution times $t$.
In particular, we observe that the second R\'enyi entropy of the total system ($i=10$) remains almost constant over time at a small value $\sim 0.4$ (corresponding to a large purity $P_2 \sim 0.8$). 
This shows that the state is slightly affected by preparation and measurement errors, but the dynamics is almost perfectly unitary. 
Considering subsystems, the second R\'enyi entropy increases as a function of time and becomes larger than the entropy of the total system, a 
conclusive signature of quantum entanglement.
Such entanglement growth has also been measured recently with superconducting qubits using RMs \cite{vovrosh2021confinement}. 

The experimental data of Ref.~\cite{brydges2019probing} have also been recently reanalyzed in Refs.~\cite{elben2020cross,elben2020mixed,vitale2021symmetry,neven2021symmetry} to access other entanglement properties, in particular using the classical shadow framework~\cite{elben2020mixed}.

\section{The many applications of Randomized Measurements}
\label{sec:oneidea}

We now turn to the myriad application of RMs. 
As described in the introduction, these applications span many areas, including: probing quantum many-body physics, quantum simulation, noise diagnostics of quantum systems, machine learning of properties of quantum systems, variational quantum algorithms and quantum computation with NISQ devices, and more. 

\subsection{Characterization of topological order}

\begin{figure*}[t]
    \centering
    \includegraphics[width=0.95\linewidth]{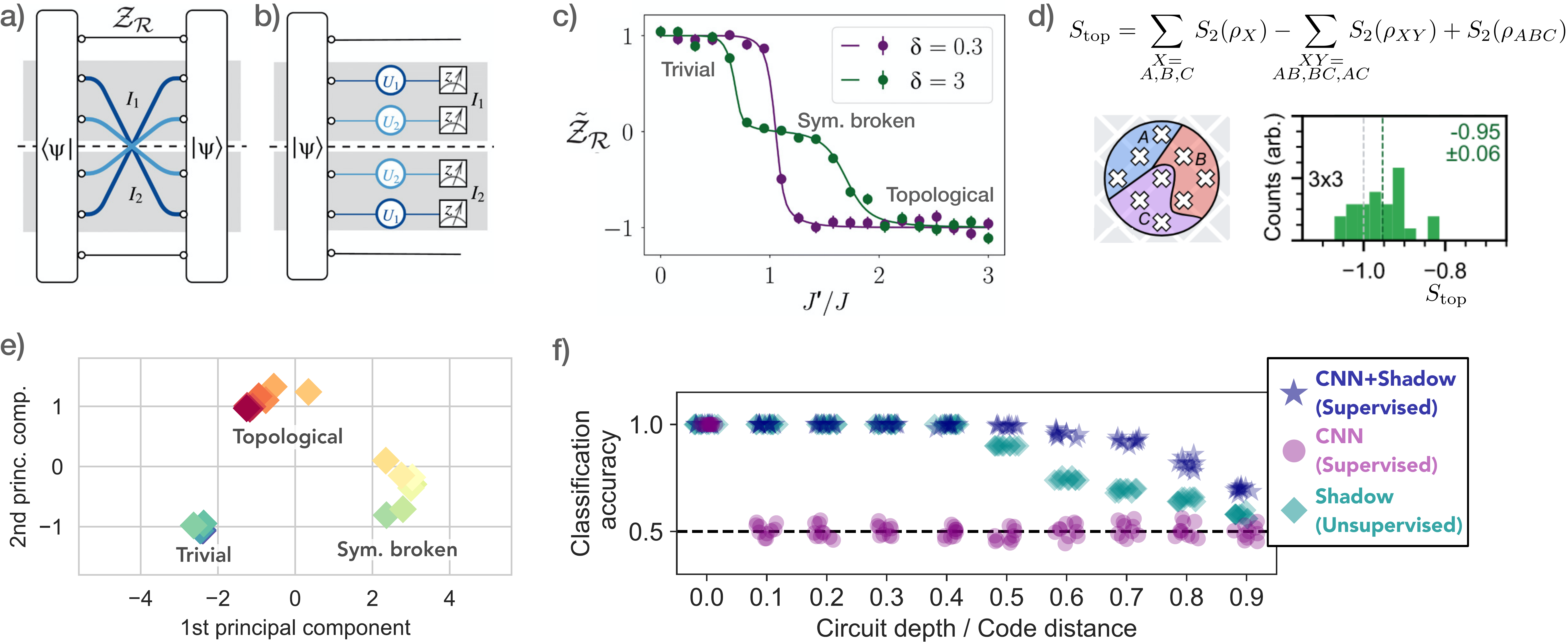}
     \caption{
    \textit{Detecting topological order with randomized measurements.}
     a,b) The many-body topological invariant ${\mathcal{Z}}_R$  associated with spatial reflection symmetry  can be inferred from statistical correlations of randomized measurements, implemented with local random unitaries applied symmetrically around the central bond  \cite{elben2020many}.
    c) Quantized values of the normalized invariant $\tilde{\mathcal{Z}}_R$ reveal two SPT phases and a symmetry broken phase in an extended Su-Schiefer-Heeger model with bond-alternating spin-exchange coefficients $J$ and $J'$ \cite{elben2020many}. 
    Dots represent estimations from  simulated RMs, lines are obtained numerically in a system with $N=48$ spins and a subsystem $I$ consisting of $n=6$ pairs of spins. 
    d)  The topological entanglement entropy $S_\mathrm{top}$ is defined from  second R\'{e}nyi entropies of various combinations of three connected partitions $A$,$B$, $C$.  The plot shows a  histogram of randomized measurement outcomes  whose average corresponds to $S_\mathrm{top}$ for the toric code ground state implemented in system of $31$ superconducting qubits \cite{satzinger2021realizing}. e) Two-dimensional feature space uncovered by an unsupervised ML model based on classical shadow \cite{huang2021provably}. Each diamond corresponds to a different quantum state in one of the three phases (trivial, symmetry broken, topological). The ML model successfully uncover the three phases by clustering them in the 2D feature space. 
    f) The accuracy (percentage of correct prediction) of different ML models for classifying between toric code phase and trivial phase under different circuit depth to perturb the states. ML models based on classical shadow (CNN+Shadow and Shadow) enjoy a much higher accuracy than CNN based on outcome from an informationally-complete POVM.}
        \label{fig:topology}
\end{figure*}

Topological quantum phases of matter are exotic phases of matter characterized by global correlations \cite{zeng2019quantum}. There is increasing interest in realizing topological quantum phases in synthetic quantum devices in the context of quantum simulation, and in topological quantum computing \cite{zeng2019quantum}. However, by their very definition topological phases cannot be detected by local measurements. Thus, their identification and  characterization in quantum simulation experiments poses a substantial challenge. RM protocols have been proposed as an experimental tool to address this challenge, and detect and classify topological quantum phases. 

 First, RM protocols have been designed to measure many-body topological invariants of symmetry-protected topological (SPT) phases \cite{elben2020many}.
 These invariants are highly non-local and/or non-linear  correlators of the many-body wavefunction.
For example, the reflection invariant, which is depicted schematically in Fig.~\ref{fig:topology}a, is written as
 $\mathcal{Z}_R=\tr{\mathcal{R}_I \rho_I }$ where $R_I$ is the 
 \textit{partial} reflection operator acting as $R_I\ket{s_1, \dots, s_{2n}}=\ket{s_{2n}, s_{2n-1} \dots, s_2, s_{1}}$  on the subsystem $I$ containing $2n$ spins symmetrically distributed across the central bond.
 If $n$ is large compared to the correlation length of the system, the quantity \mbox{$\tilde{\mathcal{Z}}_R=\mathcal{Z}_R/\sqrt{[P_2(\rho_{I_1}) +P_2(\rho_{I_2})]/2}$} acts as a topological order parameter, taking a quantized value $\pm 1$ depending on whether the phase is trivial or topological.
 (Here $I=I_1\cup I_2$, where $I_1$ denotes the $n$ spins just to the left of the central bond, and $I_2$ denotes the $n$ spins just to the right.)
 Fig.~\ref{fig:topology}b shows the RM protocol to access $\mathcal{Z}_R$ using random unitaries whose spatial distribution is reflection symmetric. As shown in Ref.~\cite{elben2020many}, the statistics of the collected bitstrings map to $\mathcal{Z}_R$.
  Fig.~\ref{fig:topology}c illustrates the protocol in the context of the (extended) Su-Schiefer-Heeger model, showing that we can distinguish a topological phase from a trivial phase. This procedure generalizes to the other topological invariants associated with time-reversal and internal symmetries,
 providing a versatile toolbox to identify SPT phases. Recently, RM protocols have been also developed to access the many-body Chern number revealing topological order in certain fractional quantum Hall states \cite{cian2020many}.

Randomized measurements were also used to identify topological order in a $31$-qubit quantum computer implementing the toric code model \cite{satzinger2021realizing}.
In this case,  RM gave access, via the measurement of the purity in connected partitions $A$, $B$, $C$ [see Fig.~\ref{fig:topology}d], to the topological entanglement entropy   $S_\mathrm{top}$ \cite{kitaev2006topological,levin2006detecting,flammia2009topological}.
The topological entanglement entropy takes a quantized value $S_\mathrm{top}=-1$ in a topologically ordered phase, and thus serves as an order parameter to detect the topological character of the toric code.
 Fig.~\ref{fig:topology}d shows results of randomized measurements for a partition of nine qubits whose average corresponds to the topological entanglement entropy. 
 
\subsection{Quantum chaos diagnostics}

Prominent diagnostics of many-body quantum chaos are out-of-time-ordered correlators (OTOCs), which detect the scrambling of quantum information
by revealing how 
local perturbations spread as a function of time \cite{swingle2018unscrambling,lewisswan2019dynamics,liu2020quantum}.
In their simplest form, OTOCs at infinite temperature  can be written as
$O(t)=\tr{ \rho_{\infty} W(t)VW(t)V }$, where $W,V$ are local operators which act on small subsystems, and 
$W(t)= e^{-\mathrm{i}Ht}W e^{\mathrm{i}Ht}$ is a time-evolved operator in the Heisenberg picture, determined by the Hamiltonian $H$, and $\rho_{\infty}\propto \mathbb{1}$ denote the maximally mixed 'infinite temperature' state.

 RMs allow to extract OTOCs at infinite temperature from statistical correlations of two separate experiments \cite{vermersch2019probing}. Importantly, no ancilla degrees of freedom and only forward time evolution are required. The idea is to generate 
in both experiments the same randomized initial state  via the application of local random unitaries $U$ to a simple computational basis state $\ket{\psi_0}$. In the first experiment, this quantum state evolves for time $t$, and we measure an expectation value $\langle W(t)\rangle_1=\braket{\psi_0|U^\dagger W(t) U |\psi_0}$; e.g.\ a Pauli operator on site $i$ is measured.  
 The second experiment is similar, except that the operator $V$, e.g.\ a Pauli operator at a different site $j$, is applied before the time evolution. We obtain then a different expectation value $\langle W(t) \rangle_2=\braket{\psi_0|U^\dagger V^\dagger W(t) VU |\psi_0}$. This is repeated for many randomized initial states, and the statistical correlations  between the two measurements $\langle W(t)\rangle_1$ and $\langle W(t)\rangle_2$ can be directly mapped to OTOCs: At initial times, the measurement of $W$ at site $i$ yields the same outcome in both experiments regardless of the application of $V$ at a different site $j$, i.e., $\langle W(t) \rangle_2=\langle W(t) \rangle_1$. These maximal statistical correlations correspond to the  maximal initial value of the OTOC. With time, the information about the application of $V$ on site $j$ spreads (``scrambles'') through the entire system, and the measurement $\langle W(t) \rangle_2$ in the second experiment  differs in general from the first measurement $\langle W(t) \rangle_1$. This decay of correlation between the measurement outcomes directly corresponds to the decay of OTOCs in scrambling quantum systems.  We note that protocol can also be extended to access finite-temperature OTOCs. In this case, one needs to sample global random states (see also Sec.~\ref{sec:beyond_local}), which are distributed according to their overlaps with respect to the thermal state of interest \cite{vermersch2019probing}.
 
Using the described protocol, infinite temperature OTOCs have been measured experimentally  in a trapped ion quantum simulator \cite{joshi2020quantum} to study the scrambling of quantum information in quantum spin models with tunable long-range interactions, and also in an NMR experiment \cite{nie2019detecting}.

Recently, such families of RM protocols based on propagating random initial states have been extended to access other quantum chaos diagnostics \cite{qi2019measuring,garcia2021quantum,joshi2022probing}.
In particular, a RM protocol has been proposed to access  the spectral form factor $K(t)=|\tr{\exp(-\mathrm{i}Ht)}|^2$ \cite{joshi2022probing} -- a quantum chaos diagnostic which is directly connected to the statics of  eigenlevels of the Hamiltonian $H$.  It can be used to test  predictions of random matrix theory and universal aspects of thermalization in many-body quantum systems. The key idea of this protocol is to apply, within a single experimental run, the same local random unitaries before and after the time evolution described by the evolution operator ${\exp(-\mathrm{i}Ht)}$.  From the statistics of final computational basis measurements one can infer $K(t)$ \cite{joshi2022probing}. Intuitively, randomized measurements implemented via this protocol allow to  sample (the product of) the traces of  $\exp(-\mathrm{i}Ht)$ and its adjoint.  
This extends the randomized measurement toolbox to access genuine properties of dynamical quantum processes, without reference to an initial state or measured observable (see also Ref.~\cite{levy2021classical} for an extension of the classical shadows framework to quantum processes). 

\subsection{Machine learning for quantum many-body problems}

By extending Theorem~\ref{thm:linear-error-bound}, one can prove that the classical data obtained from randomized measurements suffice to $\epsilon$-approximate \emph{all} reduced density matrices with a constant number of constituents in an $N$-body quantum system from only order $\log(N) / \epsilon^2$ measurements \cite[Lemma~1]{huang2021provably}.
Therefore, the succinct representations inferred from the measurement data can be used to evaluate nonlinear functions in any constant-size subsystem.

These succinct representations open new opportunities for addressing quantum problems using classical methods such as machine learning (ML). 
The overarching idea is to use quantum devices to generate many-body states with interesting properties. 
Randomized measurements allow us to convert these physical quantum states into succinct sequences of bitstrings (training data) that can then be used to train classical ML models to predict local properties or consequences thereof. 

The topological entanglement entropy is one prominent concept that fits into this framework. 
It is a nonlinear function of subsystem density operators that can be used to identify 
topological phases \cite{kitaev2006topological,levin2006detecting}.  

Importantly, it typically suffices to consider subsystems whose size is large compared to the correlation length, but 
independent of the total size of the system. A moderate number of randomized measurements 
suffice in principle for estimating this type of function. 
Alternatively, we can use randomized measurement data to train a ML model to directly classify different phases of matter, even if appropriate order parameters are not known in advance \cite{huang2020predicting}. 

Fig.~\ref{fig:topology}e and \ref{fig:topology}f illustrate the success of such an ML approach for the \emph{bond-alternating XXZ model} on $N=300$ spins (three phases: trivial, symmetry-protected topological, and symmetry broken) and the \emph{toric code topological phase} on $N=200$ qubits (two phases: trivial and topological), respectively. In both cases, the inputs are raw measurement data obtained from performing $500$ randomized measurements on an unknown state $\rho$. The trained ML model then tells us to which phase $\rho$ belongs. Indeed, Figure~\ref{fig:topology}e showcases that states belonging to the same phase cluster tightly in the feature space uncovered by the ML model.
For the toric code phase, Figure~\ref{fig:topology}f reports that ML models based on randomized measurements (CNN+Shadow and Shadow) provide accurate prediction (high on the $y$-axis) even if we perturb states in the two phases with random circuits of increasing depth ($x$-axis).

Ref.~\cite{huang2020predicting} 
provides rigorous theoretical support for these empirical studies. The proposed ML model is guaranteed to work efficiently --- i.e.\ the training data size, the number of randomized measurements, and the runtime scale polynomially in system size ---  if an underlying phase-classifying function exists. 
For the XXZ model, this function could be the many-body topological invariant from Figure~\ref{fig:topology}a. For the toric code phase, the function could be the topological entanglement entropy from Figure~\ref{fig:topology}d. Importantly, the ML model need not know this function explicitly. It is guaranteed to find it (or something even better) by itself. Hence, such ML models could be used to determine new and potentially more compact classifiers for a variety of quantum phases.

\subsection{Fidelity estimation}

\begin{figure}
    \centering
    \includegraphics[width=\linewidth]{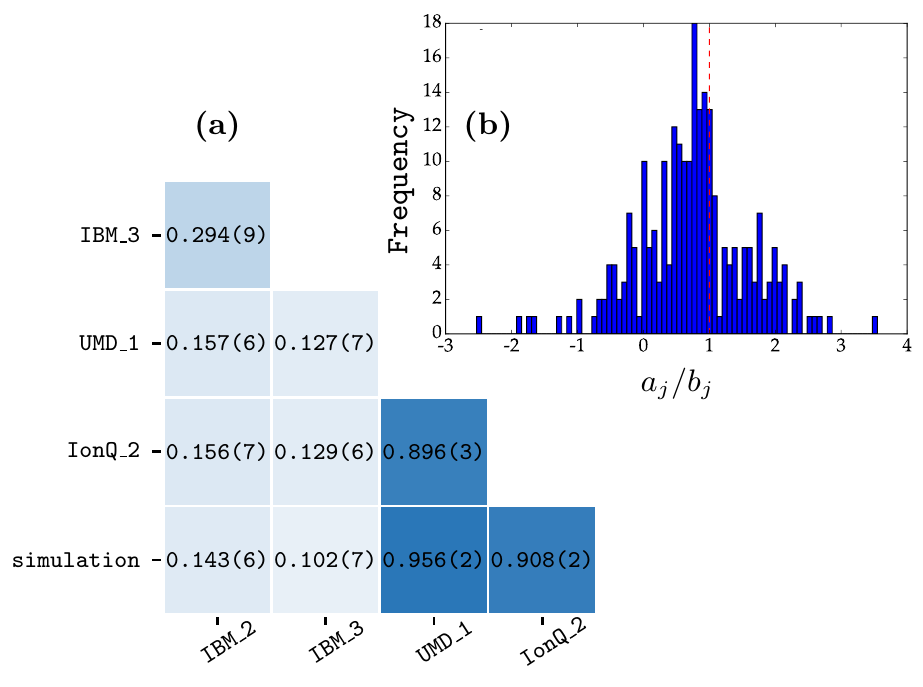}
    \caption{\textit{Fidelity estimation with randomized measurements}:  \textbf{(a)} Cross-device fidelities of 7-qubit GHZ states prepared in various quantum devices based on superconducting qubits (IBM3) and trapped ions (UMD1, IonQ2) \cite{zhu2021crossplatform}. \textbf{(b)} Direct fidelity estimation allows to extract the fidelity between an experimentally prepared quantum state $\rho$ and and a pure theoretical target states $\psi$ as the mean of the distribution of the random variable $(a_j/b_j)$ (see text). Shown is experimental data for a 14-qubit quantum state prepared in trapped ion quantum simulator with estimated mean (fidelity) of  $0.75\pm0.05$ \cite{lanyon2017efficient}.}
    \label{fig:fidelity}
\end{figure}

Suppose we wish to prepare a target state $\psi$, which we assume is pure for simplicity. 
If instead, due to experimental limitations, we can only prepare the state $\rho$, how can we tell how close $\rho$ is to $\psi$? 
As discussed above, full tomography could achieve this, but is very expensive. 
The RM toolbox provides another answer using the protocol of \textit{direct fidelity estimation} (DFE) \cite{flammia2011direct,daSilva2011practical}. 
In DFE, one first chooses an operator basis of convenient observables, e.g.~the $N$-qubit Paulis. 
The fidelity between $\psi$ and $\rho$ when $\psi$ is pure reduces to $F(\psi,\rho) = \tr{\psi \rho}$. 
Next, expand $\rho$ and $\psi$ in the Pauli basis as $\rho = \sum_j a_j W_j/2^{N/2}$ and $\psi = \sum_j b_j W_j/2^{N/2}$ for $W_j$ the $j$th $N$-qubit Pauli operator. 
Expanding the fidelity, we find that 
$$F(\psi,\rho) = \tr{\psi \rho} = \sum_j a_j b_j = \sum_j \left(\frac{a_j}{b_j}\right) b_j^2,$$ 
where the latter sum is over the support of $b_j$. 
The purpose of rewriting in this last form is that $\sum_j b_j^2 = 1$ for a pure state $\psi$, so we have reinterpreted the fidelity as an \textit{expected value} over a known distribution. 
Furthermore, the ratio $a_j/b_j = \tr{W_j \rho}/\tr{W_j \psi}$ is an observable quantity since $a_j$ can be estimated empirically and $b_j$ is known from the known target state. 
By doing Monte Carlo importance sampling of the distribution $b_j^2$ for about $O(1/\epsilon^2)$ samples, we obtain a random collection of observables $\{W_k\}$ that we can then estimate using, e.g., shadow estimation and it yields a randomized estimate of the fidelity $F$ that is accurate to within $F \pm \epsilon$ with high probability. 

Although DFE requires sampling only $O(1/\epsilon^2)$ Pauli observables, independent of $N$, there is still some important scaling with $N$ in this protocol.
First, for generic states, the choice of Paulis will include samples of very high-weight Pauli strings that must be estimated, and these are generally more difficult to estimate both as a single two-outcome measurement or as an inferred observable from many single-qubit measurements. 
Second, to estimate the fidelity of a generic state requires the ability to resolve a Pauli observable to a precision $\pm 1/2^{N/2}$. 
Thus, the worst-case complexity for the total number of measurements (not just observables) is $O(2^N/\epsilon^2)$ which is still exponential in $N$, albeit better than full tomography by another factor of $2^N$. 
However, this worst-case behavior can significantly overestimate the scaling for many important cases where most of the probability mass $b_j^2$ in the target state is concentrated on a relatively small number of low-weight Paulis. 
Important examples of such clustering behavior are stabilizer states with local stabilizer generators and high-temperature Gibbs states of local Hamiltonians. 

The above method generalizes naturally to quantum processes (see \cite{flammia2011direct,daSilva2011practical} for details), and was first used to efficiently estimate the process fidelity of a Toffoli gate in a superconducting transmon architecture~\cite{Fedorov2011implementation}. 
Another notable use of the method was to validate the fidelity of a 14-qubit state preparation \cite{lanyon2017efficient} in an ion trap, as shown in Fig.~\ref{fig:fidelity}b. 
A variant of DFE that uses a simpler Pauli measurement scheme and achieves nearly optimal sample complexity was recently proposed and tested on 4-qubit entangled states in a trapped-ion device~\cite{seshadri2021theory,seshadri2021versatile}.

While DFE allows comparison between an ideal state or process and an imperfect experimental implementation, randomized measurements also enable direct comparison between two experiments \cite{elben2020cross}. 
Here, the goal is to measure a fidelity between two mixed states $\rho_1$, $\rho_2$, defined as
\begin{equation}
    F_{\max}=\frac{\mathrm{tr}(\rho_1\rho_2)}{\max\left[\mathrm{tr}(\rho_1^2),\mathrm{tr}(\rho_2^2)\right]}.
\end{equation}
The density matrices $\rho_1$ and $\rho_2$ represent quantum states realized in different experimental devices, which may be separated by a large distance, not operating at the same time, and not using the same physical systems. Both density matrices can refer to subsystems of constant size of large quantum many-body systems.

Such a comparison between different devices is relevant in the context of verifying and benchmarking quantum computers and simulators \cite{kliesch2021theory,carrasco2021theoretical}: One might gain confidence in the result of a quantum computation or simulation by running the computation or simulation on various devices and compare, through quantitative measures such as $F_{\max}$, the outcome of one with the other \cite{elben2020cross,carrasco2021theoretical}.

The key to determining $F_{\max}$  is the measurement of the overlap  $\mathrm{tr}(\rho_1 \rho_2)$. With randomized measurements, this can be achieved as follows: (i) Generate a set of random unitaries $U$, for instance made of random single-qubit rotations (c.f.\ Sec.~\ref{sec:pro_dat}) and send them via classical communication to the two devices.
(ii) Apply these same unitaries to both $\rho_1$ and $ \rho_2$, followed by  computational basis measurements.
The overlap $\mathrm{tr}(\rho_1 \rho_2)$ can be then extracted from the statistical correlations between the outcomes obtained in both devices \cite{elben2020cross}.
One can understand this result as follows: if the two states  are identical $\rho_1=\rho_2=\rho$, the bitstrings measured in both devices will be picked from the same distribution $\braket{\mathbf{s}|U\rho U^\dag |\mathbf{s}}$, i.e.\ we will observe perfect correlations between the two experiments. 
If instead, the two states are different, the outcomes will be typically uncorrelated. Importantly, while the required number of measurements is exponential in the (sub-)system size, the cost is strongly reduced compared to performing full quantum state tomography in both devices. 
This allows to access (sub-)system sizes beyond the regime of full quantum state tomography. 

In Ref.~\cite{elben2020cross}, a proof-of-principle demonstration of this protocol was presented, by re-analyzing the data of Ref.~\cite{brydges2019probing} to compare the experimentally prepared quantum states of up to $10$ qubits with a theoretical simulation, or with a different quantum state prepared in the same experiment.
In Ref.~\cite{zhu2021crossplatform}, comparison of quantum devices across different qubit technology has been achieved by comparing entangled quantum states consisting of up to 13 qubits prepared on six different quantum devices based on trapped ions or superconducting qubits (see Fig.~\ref{fig:fidelity}a).

Let us mention that there are also approaches to specifically estimate fidelities of \emph{random quantum states}  generated by random quantum circuits. 
In this case, one can define and extract the fidelity between  prepared quantum states and their theoretical target  without needing to add another layer of randomness in the measurement stage. 
Variations of this idea are known as cross-entropy benchmarking \cite{boixocharacterizing2018,arute2019quantum,wu2021strong} and random circuit sampling \cite{bouland2018complexity,liu2021benchmarking}. 
This approach can be generalized to quantum states generated via ergodic Hamiltonian dynamics using the concept of projected state ensembles \cite{choi2021emergent,cotler2021emergent}.

\subsection{Quantum gate noise characterization}
\label{sec:gatenoise}

The previous examples have demonstrated the surprising power of random measurements, and the closely related idea of randomly applying unitary gates. 
In this section, we dive a little deeper in the direction of applying random unitaries, and we consider what happens when random unitary gates are used throughout a quantum circuit. 

Randomized dynamics were originally suggested as a way to decouple unwanted interactions from an environment and to put noise into a standard form known as a Pauli channel~\cite{Viola2005,Kern2005,Knill2005}. 
Moreover, all that is required to achieve this noise projection is the ability to insert random Pauli gates (or $\pi$-pulses) into a quantum circuit. 
A Pauli channel is any quantum channel whose Kraus operators are the Pauli matrices, and where the operator sum is weighted by a probability distribution. 
Thus, many common channels like depolarizing noise, dephasing noise, and bit-flip noise are Pauli channels, as are some more complicated correlated noise channels across multiple qubits. 
Non-examples include amplitude damping or coherent over-rotation errors. 

Pauli channels are a natural class of noise channels because their stochastic nature makes it easy to report a single figure of merit, an error rate, to a given noise process. 
They also enjoy a central role in the theory of quantum error correction because local Pauli noise can be efficiently simulated when surrounding quantum circuits are comprised solely of Clifford gates. 
Lastly, the Pauli channel error rates in a quantum device provide an important metric for progress towards fault tolerance. 

These considerations have motivated a research effort to use random unitary dynamics to simplify the noise in a quantum gate and to enable efficient characterization of noise by reducing the problem to estimating Pauli error rates. 
The literature on noise characterization is already the subject of entire review articles~\cite{kliesch2021theory,eisert2020quantum}, so we necessarily limit the scope of our discussion. 

The quintessential method for estimating average error rates in few-qubit quantum systems is called randomized benchmarking (RB)~\cite{emerson2005scalable,knill2008}.
In RB, sequences of random Clifford gates of varying length are applied to the initial state $\ket{0}^{\otimes n}$. 
At the end of the circuit, the inverse Clifford circuit is computed, compiled, and applied, then the state is measured in the computational basis. 
If the circuit had no noise, then one would always measure the $0$ outcome, but owing to noise in the system the probability of $0$ decays exponentially in the length of the circuit. 
Suppose that the noise on each Clifford gate is identical, Markovian, time-stationary noise. 
Then it can be shown~\cite{magesan2012characterizing} that the slope of this decay curve estimates the average error rate between the noise $\mathcal{E}$ and the ideal gate $U$, defined as
\[r_{\text{avg}} = 1-\int \mathrm{d}\psi \langle\psi|U^\dagger
        \mathcal{E}(|\psi\rangle\!\langle\psi|)U|\psi\rangle\,,\]
where the integral is taken over the uniform Haar measure. 
Fitting to an exponential decay by using sequences of varying lengths achieves two goals: 
first, it decouples the noise in the state preparations and measurements from the noise in the gates, improving the  accuracy of gate error estimates; 
second, the long sequences also improve the precision of the estimates by amplifying small gate errors into a signal that is observable with a reasonable amount of sampling. 
These strengths have made the RB method the \textit{de facto} standard for experimental estimation of error rates in one- or two-qubit experiments. 

The success of RB has spawned numerous modifications to improve and extend the method. 
Two early ideas in this direction were interleaved RB (IRB)~\cite{magesan2012efficient} and simultaneous RB (SRB)~\cite{gambetta2012characterization}. 
In IRB, standard RB is first performed to get a baseline average error rate estimate $r_0$. 
Then new random circuits are sampled by systematically appending to each random gate the same fixed Clifford gate $U$. 
This new experiment will generally give a worse average error rate $r$, and then the ratio $r/r_0$ provides an estimate of the average error rate of $U$. 
In this way, IRB allows one to estimate gate-specific average error rates. 
SRB works in a similar comparative manner. 
In SRB, the the baseline error rate is estimated by doing RB on a composite system, and this baseline $r$ is compared to RB done simultaneously on the constituent subsystems. 
This facilitates estimation of crosstalk error rates and correlated errors, which can be especially detrimental for fault tolerance. 
Finally, several variants have been proposed to extract just $r_{\text{avg}}$ (or related parameters) in larger-scale circuits than is possible with standard RB~\cite{proctor2019direct,proctor2021scalable}. 

As mentioned above, randomized dynamics can be used to ensure that the noise affecting a quantum computation is of the form of a Pauli channel~\cite{Viola2005,Kern2005,Knill2005}. 
This idea was further developed into a scheme called randomized compiling~\cite{Wallman2016}, which improves over naive schemes by reducing circuit depth slightly and comes with a perturbative error analysis. 
These ideas have been demonstrated experimentally in superconducting qubits~\cite{Ware2021,Hashim2021Randomized}. 

The experimental success of these methods justifies the recent effort to estimate the Pauli noise on individual gates or rounds of gates in a quantum circuit~\cite{helsen2019class,erhard2019characterizing,flammia2020efficient,harper2020efficient,harper2021fast,flammia2021pauli,wagner2021pauli,chen2021quantum,flammia2021averaged}. 
Two notable experiments in this space are an ion-trap experiment~\cite{erhard2019characterizing} that estimated the average noise on a 10-qubit M\o{}lmer-S\o{}rensen gate, and an experiment~\cite{harper2020efficient} that estimated all of the locally Clifford-averaged Pauli error rates in a 14-qubit transmon device. 
These methods have recently been put into an overarching framework called ACES (for Averaged Circuit Eigenvalue Sampling)~\cite{flammia2021averaged}. 
ACES has been shown numerically to scale to at least 100 qubits, and offers a promising avenue for scalable Pauli noise estimation in large-scale quantum devices.

\subsection{Hamiltonian \& Liouvillian learning}

Randomized measurements can be used to learn dynamical variables that govern quantum evolution such as Hamiltonians and more generally Lindbladians. 
There are many approaches to Hamiltonian and Lindbladian learning in the literature, but the quantum part of nearly all of them boils down to estimating expectation values of low-weight Pauli observables, so each of these algorithms can benefit from the RM toolbox. 

Let $H$ be an unknown Hamiltonian, and suppose one is given the ability to prepare the ground state $\psi_0$. 
Perhaps surprisingly, when $H$ is sufficiently generic, low-weight Pauli observables with respect to $\psi_0$ contain enough information to reconstruct $H$ up to an overall scale factor (and an unobservable energy shift)~\cite{garrison2018does,qi2018determining,bairey2019learning}. 
The argument is remarkably simple, and works even for a steady state $\rho$, not just the ground state $\psi_0$. 

Let us expand $H$ in the Pauli basis as $H = \sum_j c_j W_j$, and suppose that $c_j = 0$ whenever the support of $W_j$ exceeds $k$ qubits for some $k=O(1)$. 
Define the matrix $K_{lm} = i \tr{\rho [W_l,W_m]}$. 
If $\rho$ is a steady state then $[H,\rho]=0$, and furthermore for any observable $O$ we have $\tr{\rho [O,H]} = \tr{[H,\rho]O} = 0$. 
Now consider acting $K$ on the vector of Hamiltonian couplings $c$. 
By linearity, 
\begin{align}
    (Kc)_l &= i \sum_m \tr{\rho [W_l,W_m]} c_m \nonumber\\
    &= i \tr{\rho [W_l,H]} \\
    &= i \tr{[H,\rho] W_l} = 0\,.\nonumber
\end{align}
Thus, $c$ is in the kernel of $K$. 

This suggests a procedure for estimating $c$: 
First, estimate the matrix elements of $K$ by preparing the steady state $\rho$ and measuring the Pauli observables $i[W_l,W_m]$ where each $W_l$ or $W_m$ is at most $k$-body. 
If the kernel of the estimated $K$ is unique, then return a null vector as an estimate for $c$. 
(Since the kernel is only specified up to a scalar multiple, one must in general do a Rabi-type experiment to pin down the overall scale factor that completely determines $c$.) 

The precision of this estimate will depend on several factors. 
First, we must be able to prepare a steady state $\rho$. 
This can be done by time-averaging the results of the experiment~\cite{bairey2019learning,evans2019scalable}. 
Ref.~\cite{evans2019scalable} describes a protocol based on Gaussian quadrature that achieves super-exponential convergence to a steady state in the number of time averages, so this step is very efficient. 
Second, the matrix elements of $K$ must be estimated to sufficient precision. 
Notice however, that all of the matrix elements $K_{lm}$ are specified by Pauli observables of weight at most $2k-1$ (since disjoint Paulis commute) when $H$ has $k$-body couplings. 
Therefore, the $K_{lm}$ can be estimated using classical shadows specialized to Pauli observables, and it is here that randomized measurements enter. 
Finally, there must be sufficient signal in the observables so that the kernel (or approximate kernel) of $K$ is unique. 
If the (approximate) kernel is not unique, or if the least nonzero eigenvalue of $K^TK$ is small, then the problem is ill-conditioned and cannot be solved to useful precision. 

\begin{figure}
     \centering
     \includegraphics[width=\linewidth]{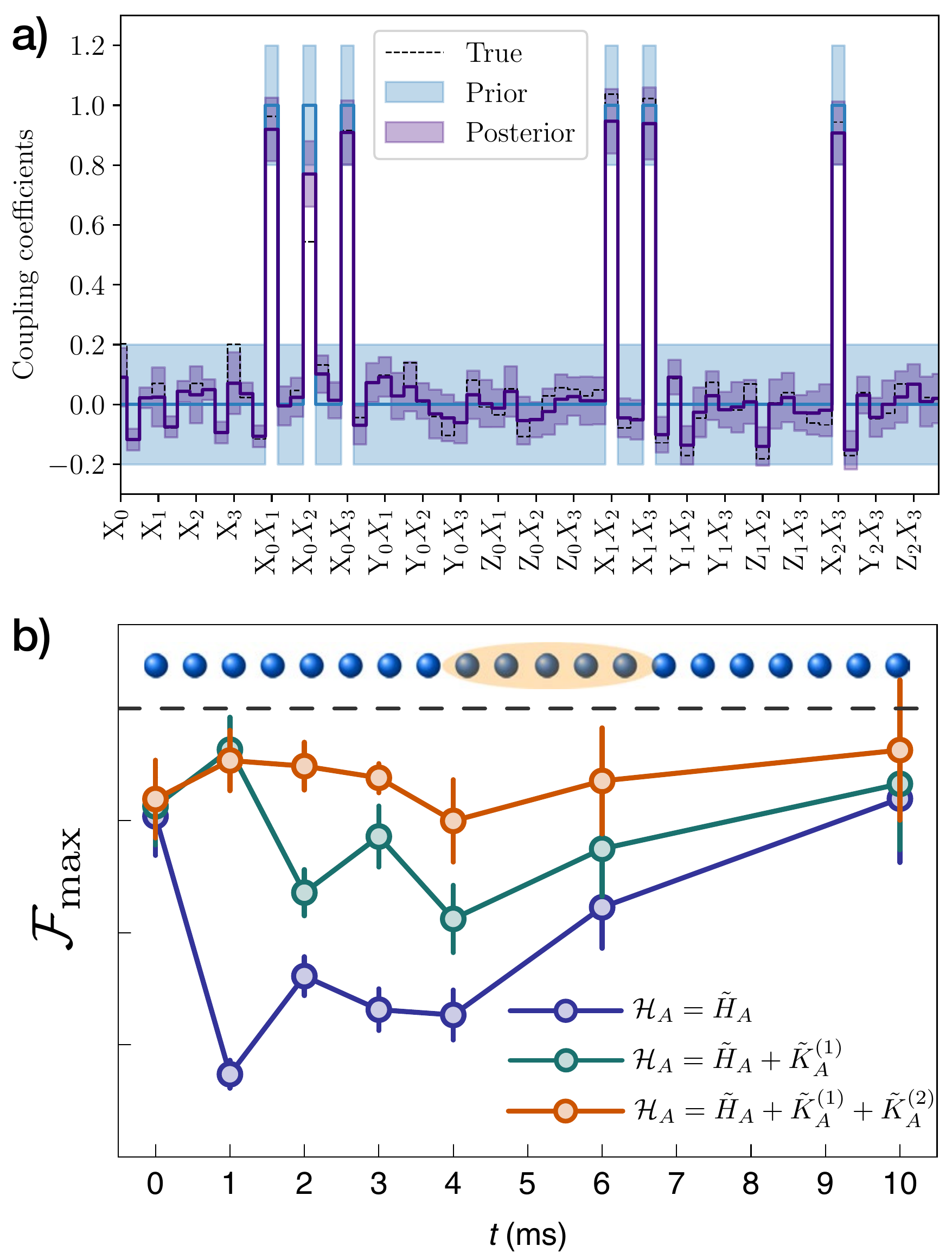}
     \caption{\textit{Hamiltonian learning with randomized measurements.} a) Scalable Bayesian Hamiltonian learning \cite{evans2019scalable} allows to efficiently estimate Hamiltonians while making use of multiple input states, well-characterized control fields or prior information on the Hamiltonian. 
     The figure shows simulation results of the true value, prior, and posterior distribution of Hamiltonian terms ($x$-axis) in a long-range quantum Ising model with $n=4$ qubits, including state preparation and measurement errors. 
     Shaded areas are drawn at two standard deviations. 
     b) Entanglement Hamiltonian tomography \cite{kokail2021entanglement} serves to efficiently learn and independently verify  the entanglement Hamiltonian $\mathcal{H}_A$, parameterizing the reduced density matrix $\rho_A\propto\exp(-\mathcal{H}_A)$ of a subsystem $A$ (here $5$ qubits). In Ref.~\cite{kokail2021entanglement} this is experimentally demonstrated using RM data collected in Ref.~\cite{brydges2019probing} in a trapped ion quantum simulator with a total number of $20$ qubits.
     Shown is the experimentally estimated fidelity $\mathcal{F}_{\max}$ of the learned  $\rho_A\propto\exp(-\mathcal{H}_A)$ at various times $t$ after a quantum quench and for different physically-motivated ansätze for $\mathcal{H}_A$. 
     At early and late times $\mathcal{H}_A$ is well approximated by a deformation $\tilde{H}_A$ of the system Hamiltonian $H_A$ (purple) while multi-body corrections $\tilde{K}^{(i)}_A$ of increasing complexity ($i=1,2$, green, orange) are important at intermediate times.}
     \label{fig:Hlearning}
 \end{figure}

This technique has been generalized in several directions. 
Ref.~\cite{evans2019scalable} has shown that Bayesian priors on the Hamiltonians can be incorporated to speed convergence, and that well-characterized auxiliary control fields can be used to enhance the precision as well. 
Fig.~\ref{fig:Hlearning}a shows numerical simulation data from this Bayesian approach applied to a long-range Ising-type Hamiltonian.
A further generalization is to replace the steady state $\rho$ by a fixed point of a Lindbladian. 
It was shown in Ref.~\cite{bairey2020learning} that this generalization admits similar guarantees as in the Hamiltonian case. 
Hamiltonian learning with this approach can also be done using the dynamics of a quenched quantum system~\cite{zhi2020hamiltonian}. 

Another avenue for generalization is provided by learning the so-called entanglement Hamiltonian~\cite{kokail2021entanglement}. 
An entanglement Hamiltonian is a Hamiltonian that describes the mixed state $\rho_A = e^{-\mathcal{H}_A}/Z$ obtained when tracing out half of a bipartite pure state $\psi_{AB}$. 
Knowing the entanglement Hamiltonian for subsystems $A$ and $B$ and their spectrum is equivalent to knowing the Schmidt decomposition of $\psi_{AB}$, and thus contains the complete information about the bipartite entanglement across this cut. 
Ref.~\cite{kokail2021entanglement} has implemented this by using randomized measurement data from Ref.~\cite{brydges2019probing}  to estimate the entanglement Hamiltonian of up to 7 qubit subsystems of a 20 qubit trapped ion quantum simulator in various states following a quantum quench. 
Figure~\ref{fig:Hlearning}b shows the experimentally estimated fidelity of the Gibbs state defined by learned entanglement Hamiltonian as a function of time after the quanum quench. 

A completely different approach to Hamiltonian learning has also recently been developed that works in a complementary regime. 
Refs.~\cite{anshu2021sample,haah2021optimal} consider learning a quantum Hamiltonian $H$ from the Gibbs state $\rho_{\beta H}$ at a small inverse-temperature $\beta$.
They showed that estimating expectations $\tr{W\rho_{\beta H}}$ of few-qubit Pauli operators $W$ enables accurate reconstruction of the Hamiltonian $H$ when $H$ couples only $k = O(1)$ qubits at a time and each qubit partakes in at most $\ell = O(1)$ interactions. 
However, it is assumed that the support of the non-zero interactions is known in advance. 
This family is a large and natural class, but some familiar systems such as two-body Hamiltonians with power-law interactions lie outside it. 
Their algorithms are unconditional in the sense that it works for any Hamiltonian in the family as long as one can prepare the Gibbs state. 
Assuming this state preparation, the estimation of $\tr{W\rho_{\beta H}}$ can be accomplished by performing randomized measurements.

While most of the work listed above focuses on qubit Hamiltonians, learning Hamiltonians of infinite-dimensional quantum systems is equally relevant. 
Ref.~\cite{hangleiter2021precise} performed Hamiltonian tomography in a nearest-neighbor transmon qubit architecture by estimating couplings between up to 6 qubits. 
They were able to synthesize these estimates across 27 total qubits to provide a comprehensive picture of the couplings in the device. 

Perhaps the most general approach to learning Hamiltonians and Lindbladians is given by notions related to gate set tomography (GST)~\cite{Blume-Kohout2016, nielsen2021gate}. 
GST grew out of the need for self-consistent estimates of a quantum process together with the noisy measurements and state preparations used as probes~\cite{merkel2013selfconsistent}. 
It works by modeling the noise on an entire collection of gates, measurements, and preparations, and then fitting all of these models to the measurements done on a large number of circuits of varying length. 
Smart choices of sub-circuits known as ``germs'' attempt to amplify small errors to make them easier to estimate. 
This basic idea has been used in dozens of experiments on 1 and 2 qubits~\cite{nielsen2021gate}, and recently it was used to characterize the 1- and 2-qubit gates in a 3-qubit experiment with two donor nuclear spins in silicon and their shared electron spin~\cite{Madzik2022Precision}. 
While GST initially focused on estimating gate sets in the form of completely positive maps, the focus broadened in experiments such as Refs.~\cite{Madzik2022Precision,samach2022lindblad} to estimating Lindbladian generators using the same or similar data fitting techniques. 

Several attempts to improve on these ideas are now being explored both theoretically and experimentally. 
On the theory side, compressive GST~\cite{brieger2021compressive} is a formulation of GST as a tensor completion problem in an effort to bring down the (substantial) computational complexity of the method and provide rigorous convergence guarantees. 
Fast Bayesian tomography (FBT)~\cite{evans2022fast} is a variant that allows the use of prior information or side information (from, e.g., randomized benchmarking experiments) to speed up GST. 
In the experiment of Ref.~\cite{evans2022fast}, the authors perform FBT on two spin qubits in silicon so quickly that the post-processing is faster than the data acquisition, demonstrating that FBT can in principle be used as an online algorithm. 
Lastly, Ref.~\cite{vandenberg2022probabilistic} learns a special case of a Lindbladian called a Pauli-Lindbladian using techniques similar to the ACES framework~\cite{flammia2021averaged} discussed in Section~\ref{sec:gatenoise} and Ref.~\cite{harper2020efficient}. 
Although this class of Lindbladians is substantially narrower than a general Lindbladian in a qubit system, it has the advantage that the learned noise can be error-mitigated very efficiently. 
This allows the authors to perform probabilistic error cancellation on a superconducting quantum device despite the presence of crosstalk errors.

\subsection{Variational quantum-classical algorithms}

Variational quantum-classical algorithms use NISQ devices as special purpose  quantum co-processors in tandem with a classical computer to solve complicated optimization problems, for instance finding the groundstate energy of a many-body Hamiltonian or variationally compressing quantum circuits \cite{cerezo2021variational}. 
Such variational hybrid approaches typically require many evaluations of complicated cost-functions through measurements on a variational quantum state. In particular, for quantum chemistry \cite{peruzzo2014variational, malley2016molecular, kandala2017ground-state} and quantum field theory applications \cite{kokail2019self}, where the cost function is represented by the energy of a complicated many-body Hamiltonian, this poses an important practical obstacle, as each cost-function evaluation requires many measurements in multiple settings. 

\begin{figure}
    \centering
    \includegraphics[width=0.48\textwidth]{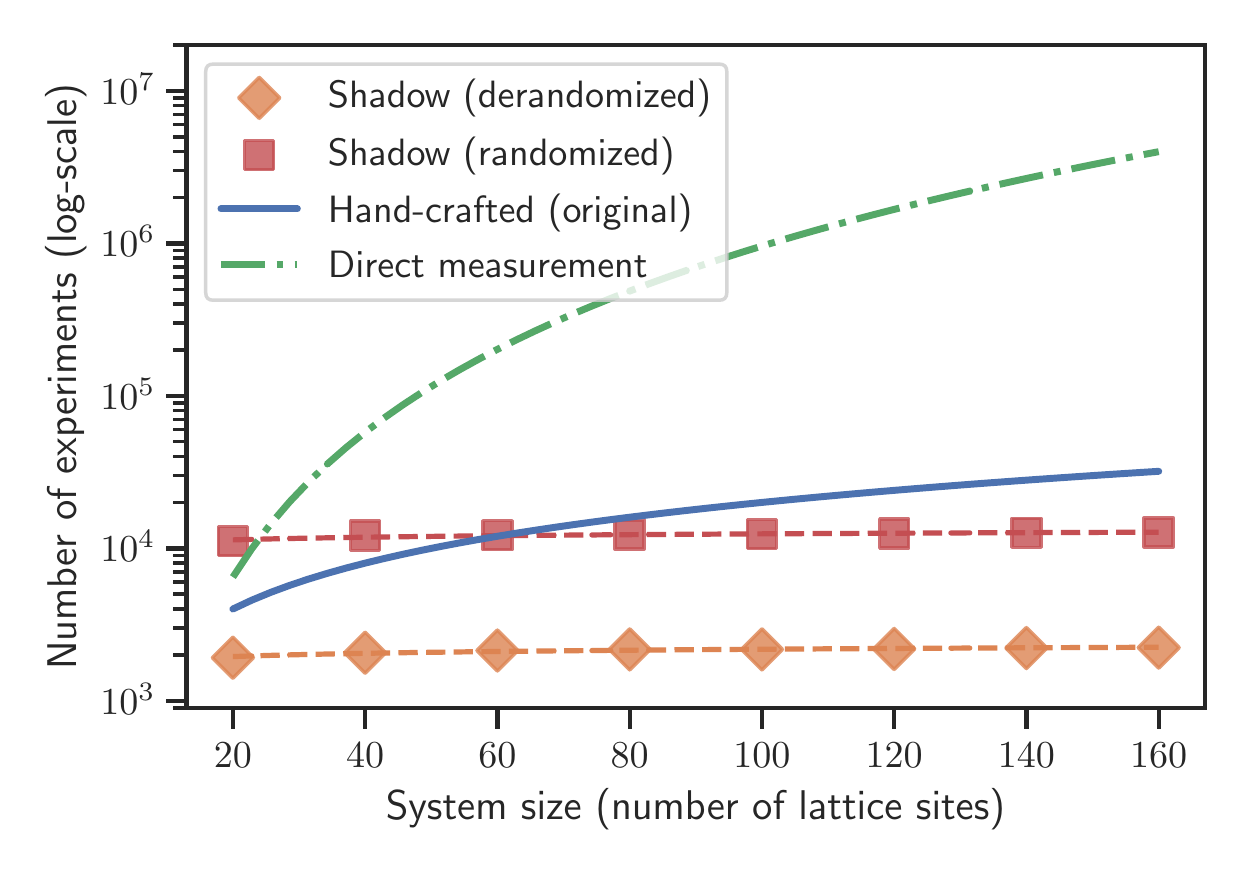}
    \caption{\textit{Variational quantum algorithm employing randomized measurements.} The number of randomized measurements  (red) to estimate the Hamiltonian variance of the Schwinger model, describing one-dimensional quantum electrodynamics, scales only logarithmically with system size, providing, for large systems, an exponential advantage over a direct measurement of Hamiltonian terms (green) or optimized 'hand-crafted' schemes (blue) \cite{kokail2019self}. This can be further improved by derandomizing the randomized protocol to find an optimized deterministic strategy \cite{huang2021efficient} (orange).}
    \label{fig:VQS}
\end{figure}

RMs can provide a big advantage here, since they allow for jointly estimating many observables (e.g.\ Hamiltonian terms) based on the same randomized measurement data via classical shadows.
According to Theorem~\ref{thm:linear-error-bound}, the required number of measurements only scales \emph{logarithmically} in the number of observables of interest  --- an exponential improvement over direct estimation protocols that estimate observables one by one. An example is shown in Fig.~\ref{fig:VQS} \cite{huang2020predicting} where the number of randomized measurements $\propto \log(L)$ required to estimate the variance of many-body Hamiltonian, involving $\propto L^2$ terms, outperforms an optimized ``hand-crafted'' scheme \cite{kokail2019self} for large system sizes $L$.

However, the required number of measurements scales exponentially with the size $w$ of the observables of interest, e.g.\ the locality of the quantum many-body Hamiltonian. This can quickly become an issue for applications 
to fermionic systems such as those encountered in quantum chemistry, where Jordan-Wigner encodings produce Hamiltonian terms with high weight  Several improvements of the elementary RM protocol (Sec.~\ref{sec:exp_rec}) are known that address this issue: \emph{Importance sampling of measurement settings} employs knowledge of the observables of interest  \cite{hadfield2020measurements,hillmich2021efficient} or of the underlying quantum many-body state \cite{rath2021importance}.   \emph{Derandomization} \cite{huang2021efficient}  replaces the randomized  measurement protocol by a deterministic one that performs, for a specific set of target observables, at least as well and sometimes much better (see also Fig.\ \ref{fig:VQS}).

\subsection{Machine learning in quantum-enhanced feature space}

In order to use NISQ devices for general machine learning problems, a class of supervised learning models using quantum-enhanced feature spaces is proposed in Refs.~\cite{havlivcek2019supervised, schuld2019quantum}.
These quantum machine learning (QML) models are trained to predict outputs, such as a real number or a discrete label, given an input vector.
The input vector, referred to as a feature vector, is transformed into a higher-dimensional quantum-enhanced feature vector using NISQ devices.
Then, the QML models train a linear function over the quantum-enhanced feature vectors via convex optimization.
Because of the convex landscape, the global optimum can always be found efficiently, without encountering any barren plateau problem \cite{mcclean2018barren}.

While training these QML models is always easy, Ref.~\cite{huang2021power} proves that the original proposal in Ref.~\cite{havlivcek2019supervised, schuld2019quantum}, which constructs quantum-enhanced feature vector based on in-place swap test, can have poor prediction performance. In simple learning tasks, Ref.~\cite{huang2021power} proves that the prediction performance can be significantly worse than classical ML models even if the QML model can perfectly fit the training data.
To resolve this issue, Ref.~\cite{huang2021power} proposes a different approach for constructing the quantum-enhanced feature vectors using randomized measurements.
The randomized measurement variant yields a simple proof of quantum speed-up in the fault-tolerant regime \cite{huang2021power}, empirically shows a significantly higher prediction accuracy over original proposals based on fidelity overlap and conventional classical ML models \cite{huang2021power}, and provides a quadratic speed-up over the original proposals in the number of quantum measurements required \cite{haug2021large}.

\subsection{Higher order polynomial functionals of the density matrix}

We have illustrated in Sec.~\ref{sec:vignette} the use of randomized measurements to measure the purity $\mathrm{tr}(\rho^2)$. Recently, expectation values of higher order $n$-copy observables of the form $\mathrm{tr}(O \rho^{\otimes n})$ have been estimated via randomized measurements. In particular, the classical shadow formalism allows us to access the expectation value of any such $n$-copy observable $O$ by cross-correlating $n$ different shadows; this extends Eq.~\eqref{eq:shadow-purity}, the special case where $n=2$ and $O$ is the swap operator, to more general $n$ and $O$ \cite{huang2020predicting}.

As an important application, randomized measurements can  be used to access moments $\tr{(\rho_{AB}^{T_A})^n}$ of the partial transpose $\rho_{AB}^{T_A}$ of a density matrix $\rho_{AB}$ describing two subsystems $A$ and $B$. These moments are expectation values $\tr{(\rho_{AB}^{T_A})^n}=\tr{\left(\Pi_A\otimes \Pi_B\right) \rho_{AB}^{\otimes n}}$ of two different $n$-copy cyclic permutation operations $\Pi_A$, $\Pi_B$ acting on $A$ and $B$, respectively \cite{elben2020mixed,zhou2020single}. There are inequalities that are satisfied by such moments if $\rho_{AB}^{T_A}$ is a nonnegative operator; thus if these inequalities are found to be violated, then $\rho_{AB}^{T_A}$ must have a negative eigenvalue, and it follows that $\rho_{AB}$ is entangled \cite{elben2020mixed,zhou2020single,neven2021symmetry,yu2021optimal}. 
In Ref.~\cite{elben2020mixed}, this was used  to experimentally demonstrate mixed-state entanglement in (sub-)systems consisting of up to $7$ qubits. 
Similarly, symmetry-resolved entropies \cite{vitale2021symmetry},  the quantum Fisher information \cite{rath2021quantum}, moments of `realigned' density matrices~\cite{liu2022detecting}, can be interpreted as expectation values of multi-copy observables  and inferred from randomized measurements. Ref.~\cite{rath2021quantum}  presents in particular closed form formulas for estimating the numbers of measurements that are required to measure an arbitrary $n$-copy observable $O$ via classical shadows.

Polynomials of the density matrix can also be used to detect and quantify multipartite entanglement. 
One way to achieve this is by studying moments of the outcomes of randomized measurements, implemented with independent local random unitary operations on each subsystem \cite{tran2015quantum,tran2016correlations,ketterer2019characterizing,knips2020multipartite,ketterer2020entanglement,imai2021bound,ketterer2021statistically}. 
Such a procedure does not require a common reference frame between the constituents, and is robust against local miscalibration of the measurement basis. 
It is thus well suited to detect multipartite entanglement in quantum networks with 
distantly separated parties \cite{ketterer2019characterizing, ketterer2020entanglement,knips2020moment}. 
Detecting multipartite entanglement in this way has been experimentally demonstrated with four-photon quantum states \cite{knips2020multipartite}. 

Finally, we note that randomized measurements recently inspired numerical sampling techniques for tensor networks which allow to estimate polynomial functionals of tensor networks states without potentially costly contraction of multiple replicas \cite{feldman2022entanglement}.

\section{Challenges and perspectives}\label{sec:challenges}

\subsection{Dealing with experimental imperfections: Robustness 
and error mitigation \rtext{Andreas, Benoit and Steve}}

In the era of noisy intermediate scale quantum (NISQ) devices, quantum operations are necessarily altered by noise and decoherence. This applies in particular to the measurement process itself.  Any practical procedure for learning properties of a quantum system must thus be equipped with sufficient robustness against noise, i.e., the ability to make accurate predictions even in the presence of (a certain level of) noise. 

Prediction procedures based on randomized measurements  involve an average over an ensemble of random unitary rotations. As such, the influence of 
noise that alters the application of the random unitary and the projective measurement can often be reduced to an averaged noise channel. 
This averaged noise channel can be efficiently learned from calibration experiments, in order to provide robust estimations.

First, estimation formulas presented in Refs.~\cite{elben2018renyi, vermersch2018unitary, elben2020cross} for purity [see Eq.~\eqref{eq:puritybitstrings}] and fidelity estimation, involve only the measured bitstrings $\mathbf{s}$. No information on the applied random unitaries $U$, other than the assertion that they are picked from an ensemble that covers the unitary group evenly (a unitary $2$-design), is required.
Thus, the estimation procedure is
insensitive to gate-independent unitary errors, e.g.\ random over- and under-rotations \cite{elben2018renyi, vermersch2018unitary, brydges2019probing, elben2020cross,vovrosh2021confinement,satzinger2021realizing}. Secondly, in the presence of well-characterized depolarization or simple qubit read-out errors, the estimation procedure can be corrected based on calibration experiments to provide unbiased estimations  \cite{vermersch2018unitary,elben2020cross,vovrosh2021confinement,satzinger2021realizing}.

In the classical shadows formalism, robust estimations can also be performed effectively in the presence of an \emph{unknown} noise channel. 
This is achieved via a calibration step that uses a state that can be prepared easily, say the state $\ket{0}^{\otimes n}$ \cite{chen2020robust,koh2020classical,berg2021modelfree}. 
By applying random unitaries that twirl the unknown noise, one can put the noise into the form of a purely stochastic Pauli channel. 
The calibration step characterizes this Pauli channel; thus the noise can be compensated in the classical postprocessing of the randomized measurement results when expectation values of observables are estimated. 
If the noise is not too strong, this procedure effectively eliminates the bias, reducing the effect of the noise on estimated observables to a floor arising from the sampling error in the characterization of the Pauli channel.

\subsection{Local vs. global random unitaries}
\label{sec:beyond_local}

 While so far we have discussed in detail implementations of RMs with \emph{local} random unitaries corresponding to single-qubit rotations, the ideas in the RM toolbox extend well beyond this. This includes, in particular,  \emph{global} random unitaries which scramble information across the entire system, implemented as quantum circuits or, in an approximate way, as random quenches. 
 
\textit{Quantum information aspects:}
This idea of using global random unitaries goes back to early work, even before their local counterparts, see e.g.\ \cite{vanenk2012measuring,ohliger2013efficient,elben2018renyi}.
Global random unitaries come with analytic expressions that are well suited for estimating certain global state properties. 
However, in contrast to local random unitaries, the type of post-processing matters a lot. 
Special-purpose formulas, like Eq.~\eqref{eq:puritybitstrings} for the purity, have global counterparts that can be computed efficiently \cite{vanenk2012measuring,elben2018renyi}. 
This, however, may not be the case for global variants of general-purpose formulas, like Eq.~\eqref{eq:classical-shadow}, because these require explicit knowledge of the random unitaries in question. This can quickly become challenging, due to the curse of dimensionality (a general unitary acting on $N$ qubits has roughly $4^{N}$ degrees of freedom). A notable exception are random multi-qubit Clifford circuits \cite{huang2020predicting,morris2019selective}. These are global scrambling unitaries that nonetheless come with efficient classical postprocessing (via the Gottesman-Knill theorem).

In the quantum circuit framework, one can also consider interpolations between local and global random unitaries. The authors of Ref.~\cite{hu2021classical} use tensor network techniques to study random unitaries based on shallow-depth circuits. In such a setup, the circuit depth provides a convenient tuning knob: Local properties are best estimated with very shallow circuits (realizing `quasi-local random unitaries'). In contrast, the number of experimental runs to estimate certain global properties  
such as the global state fidelity  decreases exponentially with the circuit depth.
A drawback of this approach is the lack of a simple analytical expression for classical shadows  and succinct property prediction  based on the obtained RM data (in contrast to RMs implemented with local random unitaries, see Eqs.~\eqref{eq:classical-shadow} and \eqref{eq:linear-prediction}).

\textit{Experimental aspects:}
The choice of implementing local vs.~global random unitaries also depends on the level of experimental control that is available in a given quantum hardware setting.
This choice can be made based on how the process of creation of these unitaries is affected by decoherence.
In an Hamiltonian spin system, RMs with global random unitaries can be implemented approximately, for example at the level of an approximate unitary $2$-design \cite{ohliger2013efficient,nakata2017efficient,elben2018renyi,vermersch2018unitary}. The idea is to use random quenches built from engineered time-dependent disorder potentials, requiring coherent interactions during a time window that increases with system size.
With a quantum computer, one can also implement random unitaries with Clifford circuits, with a size scaling quadratically with the number of qubits \cite{Koenig2014}. 
From this perspective of decoherence, local random unitaries seem thus to have an advantage over global random unitaries, as the required coherence time does not increase with system size. 
On the other hand, for implementing local random unitaries with high fidelity in e.g.\ programmable quantum simulators of trapped ions \cite{brydges2019probing} or Rydberg atoms \cite{notarnicola2021randomized}, an important assumption is that  any residual interaction between qubits can be turned off or made negligible.
Global random unitaries can be created instead in presence of an interaction `background', and may be thus seen as more appealing for certain experimental setups. 
Finally, between these two extreme cases of local-global random unitaries, the shallow-depth random circuits of Ref.~\cite{hu2021classical} can also be seen as a good compromise to realize random unitaries in an interacting system, while requiring a limited coherence time (related to the depth of the circuit). 

\subsection{Beyond qubits}\label{sec:beyond_qubits}
The RM toolbox can be extended to other quantum systems beyond qubits, in particular to systems consisting of qudits ($d$-level systems with $d\geq3$), as well as to fermionic and bosonic systems using global random unitaries. 

\textit{Qudits:} While the randomized measurement toolbox is traditionally discussed for qubits or spin-$1/2$ particles in quantum computing and quantum simulation, atomic platforms in particular offer naturally high-dimensional internal state spaces which can serve as \emph{qudits} in hardware-efficient universal quantum computing \cite{ringbauer2021universal}, and can also represent spins $S>1/2$ in quantum simulation.
We emphasize that the present protocols and applications, discussed for qubits, generalize to these cases, and random unitary operations and state-resolved measurements on single qudits are readily implementable on existing platforms of, e.g., trapped ions or Rydberg tweezer arrays \cite{monroe2021programmable,browaeys2020many}.

\textit{Fermions and Bosons:} Beyond qubits and qudits, programmable quantum many-body systems can be engineered with bosonic or fermionic particles as basic constituents. 
A seminal example is provided by ultracold fermionic atoms in optical lattices described by a 2D Fermi Hubbard model with repulsive interactions, where state of the art experiments achieve single-site control and site-resolved single-shot readout via a quantum gas microscope  \cite{NAP25613}. 
These setups provide the toolbox to prepare and study strongly correlated equilibrium phases and non-equilibrium dynamics. 
We note that {\em local} random unitaries as discussed above are typically not available in these experiments, as physical Hamiltonians generating unitaries are constrained by conservation laws, such as atom number number in closed systems. Instead a RM toolbox can be developed based on {\em global} random unitaries, which have a block structure inherited from the conservation of particles \cite{ohliger2013efficient,elben2018renyi,vermersch2018unitary}.

\subsection{ Learning about the quantum 
world using classical machines \rtext{Richard and Robert}}

In order for a classical machine to learn, store, and manipulate any object of interest, we must construct a classical representation of such an object.
Randomized measurements provide a powerful set of tools for converting quantum systems into efficient classical representations that capture many aspects of the original quantum object.
These tools bridge the gap between the quantum and classical worlds.
Any algorithm originally designed for the purpose of learning in a classical world can now be used to learn about a quantum-mechanical world by employing randomized measurements as a quantum-to-classical converter.

Recently developed classical algorithms for learning in a classical world are capable of predicting what would happen in scenarios never encountered before \cite{lecun1998gradient, goodfellow2016deep, Silver2017MasteringTG, jumper2021highly}.
Some well-known examples include outperforming the best human players in games, answering questions after reading an article, and identifying potential illnesses in the human body.
By combining these classical algorithms with quantum-to-classical converters, we envision that classical machines may one day achieve a powerful ability to predict the behavior of the quantum world as well.
The potential applications range from predicting properties of exotic quantum systems that have not previously been realized in the physical lab \cite{huang2021provably}, to designing better quantum computers, to 
discovering new physical phenomena. Thus we anticipate that quantum-to-classical conversion enabled by the randomized measurement toolbox will have a vital role in unraveling some of nature's deepest secrets.

\begin{acknowledgments}
B.V.\ acknowledges funding from the French National Research Agency (ANR-20-CE47-0005, JCJC project QRand), and from the Austrian Science Foundation (FWF, P 32597 N).  	A.E.\ acknowledges funding by the German National Academy of Sciences Leopoldina under the grant number LPDS 2021-02 and by the Walter Burke Institute for Theoretical Physics at Caltech.
Work at Innsbruck is supported by
the US Air Force Office of Scientific Research (AFOSR)
via IOE Grant No. FA9550-19-1-7044 LASCEM, the European Union’s Horizon 2020 research and innovation program under Grant Agreement No. 817482 (PASQuanS),
and by the Simons Collaboration on Ultra-Quantum Matter, which is a grant from the Simons Foundation (651440,
P.Z.). 
J.P.\ acknowledges funding from  the U.S. Department of Energy Office of Science, Office of Advanced Scientific Computing Research, (DE-NA0003525, DE-SC0020290), and the National Science Foundation (PHY-1733907). The Institute for Quantum Information and Matter is an NSF Physics Frontiers Center. 
\end{acknowledgments}

%


\end{document}